\documentclass{article}

\PassOptionsToPackage{numbers,sort&compress}{natbib}
\usepackage[sglblindworkshop, final]{neurips_2026}

\usepackage[utf8]{inputenc}
\usepackage[T1]{fontenc}
\usepackage{hyperref}
\usepackage{url}
\usepackage{booktabs}
\usepackage{amsmath,amsfonts}
\usepackage{nicefrac}
\usepackage{microtype}
\usepackage{xcolor}
\usepackage[most]{tcolorbox}
\usepackage{graphicx}
\usepackage{caption}
\usepackage{subcaption}
\usepackage{titlesec}
\usepackage{enumitem}
\usepackage{wrapfig}

\setlist{nosep}

\title{An Investigation of the NeurIPS and ICML 2025 Position Tracks}

\author{
    Fan Yang$^{1}$, Wenkai Li$^{2}$, Jun Liu$^{2}$\\
    $^{1}$Fujitsu Research, $^{2}$Carnegie Mellon University
}

\begin{document}
\maketitle

\begin{abstract}
ML venues shape what kinds of research claims become legible to reviewers and what forms of evidence count as rigorous. The NeurIPS and ICML Position Paper Tracks were created for agenda-setting work, making their early composition worth auditing. \textbf{This paper argues that the publicly accessible 2025 reviewed pool is dominated by reformist critique, and that the track should explicitly solicit direction-setting work alongside, not in place of, the reformist critiques it already hosts well.} We audit every accessible submission to the NeurIPS 2025 and ICML 2025 Position Tracks under a pre-specified rubric, and compare the resulting pattern with a reference class of widely recognized agenda-shifting ML papers. Three-quarters of audited submissions critique an existing benchmark, evaluation, or methodology; these papers score highly on our artifact-coupling rubric, but evidentiary depth does not predict reviewer rating. The reference class (AlexNet, the Transformer, Concrete Problems in AI Safety, and others) differs from the accessible reviewed pool in \emph{artifact kind}: agenda-shifting papers typically gave the field something new to build on, test against, or contest, such as a measurement protocol, benchmark proposal, toy implementation, dataset card, audit template, or falsifiable experimental program. We close with four CFP-level interventions aimed at broadening the submission mix without displacing the critiques the track already hosts well.
\end{abstract}

\section{Introduction}
\label{sec:intro}

Research agendas rarely shift through argument alone. They shift when a claim gives others something to build on, test against, or contest: a measurement protocol, a benchmark proposal, a toy implementation, a dataset card, an audit template, or a falsifiable experimental program. Such artifacts need not be large systems or released models. Their function is to make a position operational---to turn a perspective into a tractable object for subsequent work.

The NeurIPS Position Paper Track, like its ICML counterpart, was created to host papers that ``shape future scholarly dialogue'' by ``advancing a new perspective, idea, or research direction.'' This makes the track an important venue-design experiment. In its formative years, the papers that become visible through review do more than populate proceedings; they teach future authors what a position paper is expected to look like, what kinds of evidence appear reviewable, and which forms of ambition can be made credible within the conference review process.

We audit the publicly accessible reviewed pool from the NeurIPS 2025 and ICML 2025 Position Paper Tracks ($N = 191$; ICML-to-NeurIPS resubmissions count as separate papers;~
\citep{%
  or_1RlrtH6ydW, or_1ZC4RNjqzU, or_1rh8iTehBc, or_3wEY0NB2pG, or_42Au7FoD8F, or_4KhDd0Ozqe,
  or_4UhTWPwVke, or_5Hpm74b1Ga, or_8samCaCwKu, or_9skHxuHyM4, or_A0HtZM0MpZ, or_ACzL62Jp4E,
  or_Al5mEX6eHF, or_BCP8UU2BcU, or_CA9NxmmUG5, or_CYJlJgEzZs, or_DMRrbb36r5, or_ET6qJpllEi,
  or_ErKu9lP91g, or_GYZLed4d3M, or_GrBXso0e17, or_H72JEXAPwo, or_HuvAM5x2xG, or_IxCvgUme5S,
  or_J5MmGPWKfb, or_JMoWFkwnvv, or_JkcSsFWdGP, or_L6RpQ1h4Nx, or_LEYmr1TsBW, or_LL39y0Tfxb,
  or_LkdH35003E, or_Lrv20S5RZV, or_MxCJbuJhWG, or_Nr2ulBN50q, or_QLKBm1PaCU, or_QMgCDPWL9Y,
  or_RrvhbxO2hd, or_RuLsq4LSZK, or_Rxd2TpV6Eg, or_TEkyydR6il, or_UTxi86wmas, or_UXZJ3aL8vE,
  or_V1FP9WDKa7, or_WpePuya3Ki, or_YhZ2PY2nZa, or_YmTxiR1HUX, or_YuMEUNNpeb, or_asQJx56NqB,
  or_b2gM1HyAgE, or_bXfF6Dqe9s, or_cRBg1dtj7o, or_cw7MYyDL33, or_eI8KegpPyX, or_eax2ixyeQL,
  or_fRk0nKLKrJ, or_gCPJFcHskT, or_gnyqRarPzW, or_gwhPvu97Gm, or_hxwIndG0Z8, or_j3totqf8xW,
  or_j5eEF77JQz, or_jd1N60VNFE, or_kJzB6lQmcb, or_l8QemUZaIA, or_mWlnrtOTtm, or_mzc1KPkIMJ,
  or_nDFpl2lhoH, or_nrlGUdlo16, or_q7QJyxgAq4, or_tctWi7I5wd, or_wkisIZbntD, or_yYJo8czj4f,
  or_zfohnbkMu0, or_1AS7IT3X2A, or_2qErp0n78A, or_3tdITbjv8t, or_75WZP8whT8, or_ASOb5Cw8Rv,
  or_DH84SmPVxR, or_ELErARGR5U, or_GCqffUAyiu, or_Gl3V36BkCL, or_I7nESnBvib, or_KL8bi77TFF,
  or_Njaljr4V8L, or_OT3WJRwCIf, or_QlEfwyq9rx, or_SlgXCLZFj3, or_Sz90WdONPz, or_T21ixFVruG,
  or_Uh13utzs2m, or_d2k8uDYlkh, or_jK4dbpEEMo, or_n1rqG1LnRF, or_vMTijVnXQ8, or_wjQqi7DJM2,
  or_0ngi2StMwC, or_1IpHkK5Q8F, or_6plSmhBI33, or_BzFMBNqg7R, or_DS1XSAPvKs, or_Ev5xwr3vWh,
  or_EvXWexakZX, or_FfsxgSZW0c, or_FjxyAotxtT, or_HzGZVYi8fK, or_OMc0BYxND4, or_PFRandBfSz,
  or_PegEYWWXvx, or_PgA9rZoMY8, or_R5uuqCAPf8, or_RT3Jby7v21, or_SbfjBNlJE7, or_USqNoPVhxx,
  or_V6DcL5L6CU, or_VZnOKzQ5qW, or_a4oXTW1PW2, or_aVwhTcSMl4, or_d7hqAhLvWG, or_dl5pvd5IgW,
  or_fXiPp3qvrW, or_gXTFLbGUQp, or_mXBFoHDuil, or_mdKzkjY1dM, or_mfd6GRW4Az, or_qh9eGtMG4H,
  or_rdeCalg68L, or_tp94g4Vmad, or_uEY7kQsiZz, or_upugtLPOxC, or_vFae5rRman, or_vlM9rLv5xB,
  or_xcdlSMYXxD, or_xnNHXepQ9h, or_yZU8kdwafM, or_yqKfMr0yvY, or_0TRVB5ghCR, or_4lyIMYzILk,
  or_5X4GDSUumr, or_816gaVGHgP, or_8Ow7kh78fk, or_8ZH52QHIZV, or_Aa50oIovvD, or_AsC0NOkJ2m,
  or_BXLRMWLDQw, or_EIEVNiraPS, or_FAD6MEhQQR, or_FJF1sa6elQ, or_JCqGIUAsbs, or_LAXgS0xzPf,
  or_NHDOjeVMb5, or_NILrMDAqEt, or_NtJfzzleG8, or_Omq9tUouSS, or_Pcys8py9RL, or_R6TXwNF1SB,
  or_RV12OsgCO0, or_RZRRb11jXp, or_RkX3UyGunC, or_RyBZXCVr1k, or_TDZjksboWO, or_U46jD48SJi,
  or_V5PNJ5HnpA, or_Vib3KtwoWs, or_XR9UpqWhmT, or_YQplP7XrAo, or_Z6UueXyYwB, or_ZOUHFrCmwu,
  or_a9eBWrd5Jg, or_aXMPvmBAm5, or_bEhRgt7mwt, or_cIbQaSXqYm, or_dVKcLgcCLZ, or_g8Fo6qtnMR,
  or_gIIqPel6w5, or_gY0BOsPO0k, or_iBkQYeEfzn, or_iOSHFKHQNP, or_j0h4glzL2F, or_j5Qmcv9jtc,
  or_kEdP6usKZd, or_kJfpS7lCVT, or_nKpmLCN0Q9, or_o3M9ibtZWV, or_oz2QmdrPdz, or_pRiGl7qF0v,
  or_tMJvb9JDsd, or_u0FB996GIH, or_uoGQOg1oxZ, or_yZhVKDW0o0, or_ygfzWIGDN8%
}). Using a pre-specified rubric, we classify each paper by claim type, artifact relation, falsifiability, and evidentiary structure. Our goal is not to infer the full distribution of submissions, since withdrawn and desk-rejected papers are not public. Instead, we ask what kind of position paper is visible in the reviewed pool that authors and reviewers can observe.

The answer is striking: the publicly accessible 2025 reviewed pool is dominated by reformist critique. Most papers propose adjustments to an existing benchmark, evaluation, dataset, or methodology rather than operationalizing a new research direction. Many of these critiques are rigorous and valuable. The concern is not that the track should host fewer of them, but that a venue created for direction-setting work may, by default, make reformist critique the most legible and reviewable form of position-taking.

\textbf{We suggest that the track should explicitly solicit direction-setting work—especially papers that pair a position with an operationalizing artifact—alongside, not in place of, the reformist critiques it already hosts well.} Three observations support this claim. First, roughly three-quarters of audited papers are reformist critiques. Second, the evidentiary depth captured by our rubric does not reliably predict reviewer score. Third, compared with a historical reference class of agenda-shifting ML work, the accessible reviewed pool is concentrated less around artifacts that open new lines of work than around critiques of artifacts that already exist. The issue is therefore not quality but balance: the visible pool contains comparatively few papers that make a new direction concrete enough for others to pursue.

\section{An Audit of NeurIPS 2025 and ICML 2025 Position Papers}
\label{sec:audit}

\subsection{Dataset and method}

We pulled all accessible submissions to the NeurIPS 2025 and ICML 2025 Position Paper Tracks from OpenReview. After filtering for papers with retrievable PDFs and at least one complete review, we obtained a pooled corpus of $N = 191$ papers: 95 from NeurIPS (40 accepted, 55 rejected) and 96 from ICML (73 accepted, 23 rejected).\footnote{Rejected submissions withdrawn by their authors, and desk-rejections, are not public on OpenReview, so our ``rejected'' count under-reports the true rejection pool; the within-pool accept/reject proportions are not venue acceptance rates and may reflect visibility policy rather than selectivity. Withdrawal may also be missing-not-at-random with respect to claim type, biasing within-pool accept/reject comparisons. We treat acceptance-based results as robustness checks, not primary evidence.} The analyses in this paper are conducted on this accessible pool; we discuss what this sampling implies for interpretation in \S\ref{sec:limitations}.

For each paper, we used a structured-output LLM classifier (Google Gemini 3 Pro, 3 independent runs per paper, consolidated by majority vote) to extract eight fields from the full paper text: a one-sentence thesis summary, a four-category claim-type label, an artifact-coupling score on a 0--5 rubric, a six-category artifact-type label, two falsifiability indicators (one for an explicit statement, one for an inferable claim), an audience label, and a citation-evidence flag. 
The four claim-type categories are defined briefly here, with the distinction being the \emph{primary} orientation of the paper:

\begin{itemize}[leftmargin=1.4em, topsep=1pt, itemsep=0pt]
\item \textbf{Normative}: proposes that the community should adopt a new direction, task, or commitment not currently pursued. Example prototype: ``ML safety should be a mainstream research agenda.''
\item \textbf{Reformist}: proposes that the community should adjust or improve an existing practice. Example prototype: ``Benchmark $X$ is saturated; we should evaluate differently.''
\item \textbf{Descriptive}: characterizes the state of the field without a clear ought-claim.
\item \textbf{Predictive}: makes falsifiable claims about where the field is heading or what will become important.
\end{itemize}

We validated classifications through a re-audit using Claude Sonnet 4.6, obtaining per-field agreement ranging from 0.50 (the artifact-coupling score, our most subjective rubric field) to 1.0 (presence of an explicit falsifiability statement, a near-absent feature in the corpus). For claim type, the field central to our headline findings, cross-model agreement was 0.85. The Gemini 3-run self-consistency on claim type was 0.90; on artifact-coupling, 0.85. 
Higher-agreement fields anchor our primary findings; lower-agreement ones, secondary.

To address the anchoring caveat on the cross-model figure, we conducted a blinded expert re-audit of \texttt{claim\_type}: a single human annotator, shown the full text of each paper, independently assigned the primary claim-type label while blinded to all model outputs. Raw agreement with Gemini's majority vote was 0.81: slightly below the 0.85 cross-model figure, in the direction the anchoring would predict, and compatible with the F1 robustness claim below. A single annotator cannot separate rubric ambiguity from individual idiosyncrasy, and raw agreement is not chance-corrected (the 74.9\% reformist prior inflates it relative to a Cohen 's-kappa-style measure). The 0.81 figure complements the 0.85 cross-model and 0.90 Gemini self-consistency numbers under a cleaner blind protocol, rather than serving as an independent gold standard.

\paragraph{Reviewer rating scales and pooled normalization.} NeurIPS 2025 reviewers use a 1--10 rating scale; ICML 2025 reviewers use a 1--5 scale. Pooling raw ratings across the two venues is not scale-comparable: a median-rated ICML paper (rating 3 on 1--5) appears below a near-minimum NeurIPS paper (rating 3 on 1--10) when the two columns are concatenated. For any pooled rating-based analysis in this paper we rescale ICML ratings by $\times 2$ to place both venues on the same 1--10 range, and annotate all rating-based figures accordingly. Per-venue analyses are unaffected (the rescaling is strictly monotonic within a venue, so within-venue rank statistics are invariant). Rank-based pooled tests that depend only on inter-venue rank order are sensitive to this choice: we report them under the normalized scale and flag where the raw-pooled number would have differed. Linear $\times 2$ assumes proportional calibration between the two scales, which we cannot verify from rating distributions alone. The findings that carry the paper's argument do not depend on this choice: F1 does not use ratings; F3's nulls hold within each venue where the rescaling is rank-invariant; and the artifact-kind contrast in \S\ref{sec:history} is a statement about \texttt{artifact\_type}, not ratings.

\subsection{Findings}

\begin{wraptable}{r}{0.36\linewidth}
\centering\small
\vspace{-1.1em}
\caption{Claim-type composition of the audited pool ($N = 191$).}
\label{tab:claimtype-composition}
\begin{tabular}{lrr}
\toprule
Claim type  & $n$ & \% \\
\midrule
reformist   & 143 & 74.9 \\
normative   & 33  & 17.3 \\
descriptive & 11  & 5.8  \\
predictive  & 4   & 2.1  \\
\bottomrule
\end{tabular}
\vspace{-1.2em}
\end{wraptable}

\paragraph{F1: The corpus is dominated by reformist papers.} Of 191 submissions, 143 (74.9\%) are classified as reformist, 33 (17.3\%) as normative, 11 (5.8\%) as descriptive, and 4 (2.1\%) as predictive (Table~\ref{tab:claimtype-composition}). The reformist plurality holds in both venues (70.5\% at NeurIPS, 79.2\% at ICML), with normative somewhat more common at NeurIPS (23.2\% vs.\ 11.5\%). The plurality remains above 60\% even under pessimistic reassignment of the 15\% cross-model disagreement rate on claim-type labels, so the qualitative pattern is robust to classifier noise.

\paragraph{F2: Papers are already well-coupled to artifacts.} The mean artifact-coupling score is 3.08/5 (SD 1.19); the modal value is 4 (43\% of papers). Only 12.6\% of papers have no identifiable artifact. 46\% present a measurement study; 23\% critique an existing benchmark; 17\% propose a concrete experiment. The issue is therefore less the absence of empirical material than the kind of artifact being centered.

\begin{figure}[t]
\centering
\begin{subfigure}[t]{0.38\linewidth}
\centering
\includegraphics[width=\linewidth]{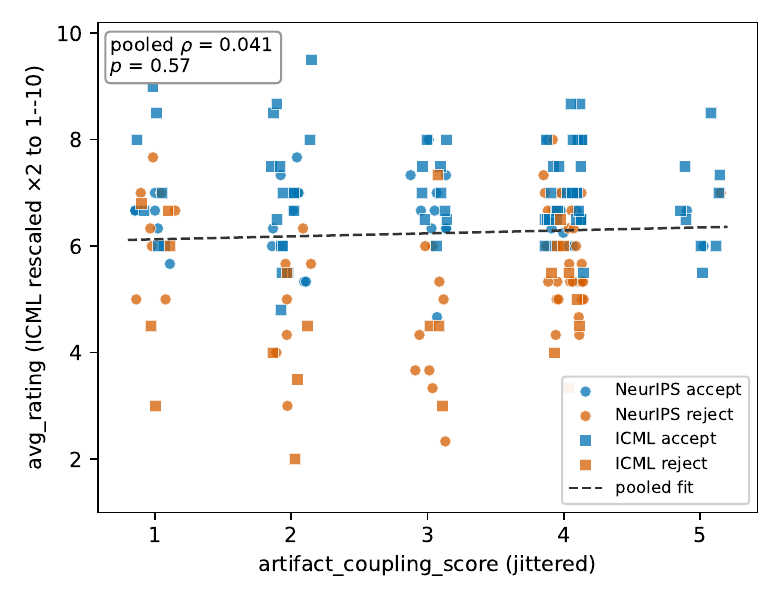}
\caption{Reviewer average rating vs.\ LLM-scored artifact-coupling, all 191 papers (x-axis jittered; ICML ratings rescaled $\times 2$).}
\label{fig:audit-scatter}
\end{subfigure}\hfill
\begin{subfigure}[t]{0.6\linewidth}
\centering
\includegraphics[width=\linewidth]{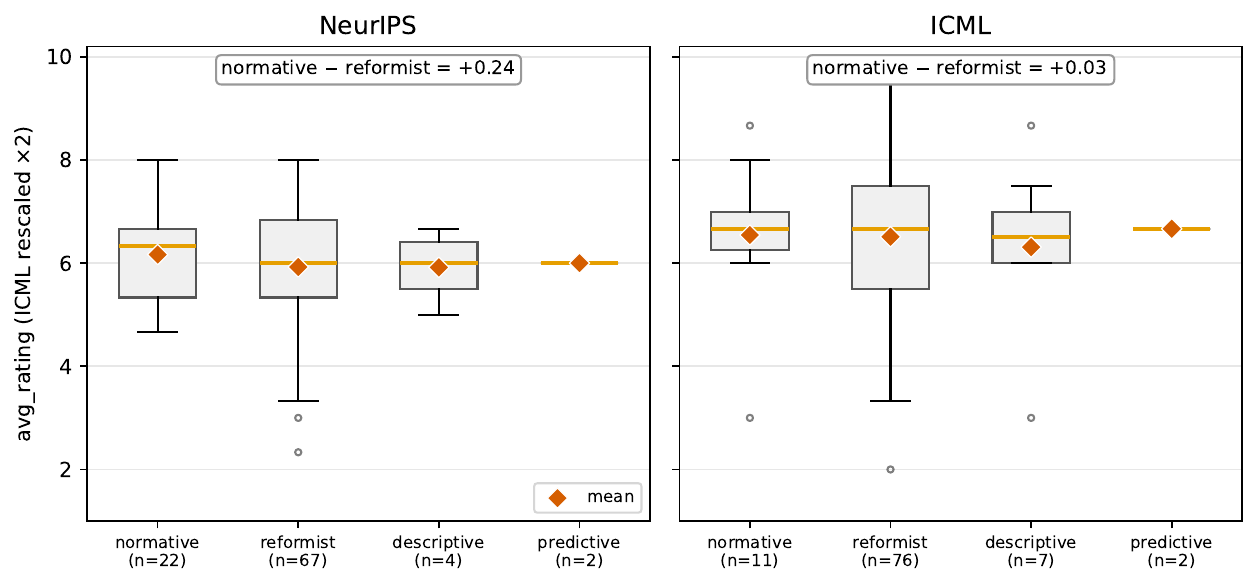}
\caption{Reviewer rating by claim type, per venue (ICML rescaled $\times 2$). Vermilion diamonds: means. Orange lines: medians. Statistics and sample sizes in the main text.}
\label{fig:claim-type}
\end{subfigure}
\caption{(a) Artifact coupling does not predict review outcome. (b) Within each venue, normative papers rate slightly above reformist, with small and venue-dependent magnitudes.}
\label{fig:audit}
\end{figure}

\paragraph{F3: Artifact coupling does not predict reviewer score.}
Figure~\ref{fig:audit}(a) plots reviewer score against artifact-coupling score, separated by venue and decision. Across the normalized pooled sample, the association is essentially zero (Spearman $\rho=+0.042$, $p=0.57$). The same pattern holds within venues: $\rho=-0.028$ for NeurIPS and $\rho=+0.067$ for ICML. Accepted and rejected papers also show substantial overlap in artifact-coupling scores within each venue.
This does not show that reviewers ignore evidence. It shows a narrower point: under our rubric, papers with deeper or more explicit artifact coupling do not receive reliably higher reviewer scores in the accessible reviewed pool.

\paragraph{F4: Normative papers receive at most a weak ratings advantage.}
Figure~\ref{fig:audit}(b) compares reviewer ratings by claim type within each venue, with ICML ratings rescaled to the NeurIPS 1--10 scale. Normative papers score slightly above reformist papers in both venues, but the gaps are small: $+0.25$ points at NeurIPS and $+0.04$ points at ICML after normalization. This pattern suggests, at most, a weak venue-level preference for normative ambition.

The pooled evidence does not support a stronger conclusion. After normalization, normative and reformist papers have nearly identical mean ratings, 6.29 versus 6.24, and nonparametric tests show no meaningful separation either across claim types or between these two categories specifically (Kruskal--Wallis $H=0.19$, $p=0.98$; Mann--Whitney $U=2354.5$, $p=0.99$, $r_{\text{rb}}=0.002$). A raw pooled test would misleadingly appear significant because it mixes incompatible rating scales: normative papers are disproportionately from NeurIPS ($22/33=67\%$), whose native scale is twice ICML's. Once ICML scores are rescaled, the apparent pooled advantage disappears.

We therefore treat F4 as a transparency finding rather than as evidence for the paper's position. The direction of the effect is consistent across venues, but the magnitudes are small, the result is post-hoc, and the apparent pooled significance depends on an invalid scale mixture. The main argument rests instead on the compositional imbalance in F1, the artifact-kind contrast in F3, and the historical comparison in \S\ref{sec:history}.

\paragraph{F5: Falsifiability is asymmetric between stated and inferred.} Only 1 paper out of 191 (0.5\%) contains an explicit falsifiability statement: a sentence of the form ``this position would be undermined by the following observation.'' By contrast, under our permissive inferred-falsifiability field, the classifier judged \emph{every} paper to contain an inferable falsifiability claim. The gap is revealing: the reviewed pool contains many claims that can be made testable, but almost none state the test explicitly.

\section{Historical Anchor: A Narrative Contrast}
\label{sec:history}

\begingroup
\setlength{\parskip}{0pt plus 0.3pt}
\titlespacing*{\paragraph}{0pt}{0.2ex plus 0.1ex minus 0.1ex}{0.5em}

To interpret the accessible 2025 pool, it is useful to contrast it with a small set of widely recognized agenda-shifting ML papers from 2012--2022. We do not treat these as a statistically representative sample, nor do we claim that agenda shifts must take this form. The list is qualitative: it supplies a reference point for the \emph{kinds of artifacts} that have anchored new research directions.

\paragraph{AlexNet (2012).} \citet{krizhevsky2012imagenet} demonstrated deep convolutional networks on ImageNet \citep{deng2009imagenet}, triggering the vision community's pivot to deep learning. The paper itself is an engineering paper, not a position paper. But its \emph{effect} on agendas was normative-by-demonstration: ``you should be training deep networks now.''

\paragraph{The Transformer (2017).} \citet{vaswani2017attention} displaced recurrence as the dominant architecture for sequence modeling. Again, an engineering paper whose normative effect was delivered through capability rather than argument.

\paragraph{Adversarial examples (2014).} \citet{szegedy2014intriguing} and \citet{goodfellow2015explaining} crystallized the adversarial-robustness subfield, arguing via demonstration that this class of failure would matter more over time. Subsequent research confirmed the prediction.

\paragraph{Concrete Problems in AI Safety (2016).} \citet{amodei2016concrete} is the closest thing in the reference class to a pure position paper: a prose taxonomy of five safety problem categories, each with proposed experimental directions that turned out to structure much of the subsequent alignment literature.

\paragraph{Stochastic Parrots (2021).} \citet{bender2021stochastic} combined warnings about LLM failure modes with normative claims about research allocation, giving later work a shared vocabulary for contesting the costs of scale.

\paragraph{The ``bitter lesson'' (2019).} \citet{sutton2019bitter} is a short essay codifying a pattern of AI progress, with a normative corollary. It functions as a Schelling point for subsequent discussion.

\paragraph{RLHF's rise to mainstream alignment (2017--2022).} \citet{christiano2017deep} was a capability demonstration; \citet{ouyang2022training} was the deployment-scale demonstration that mainstreamed the approach. Neither is a position paper; both deliver normative weight through engineering.

\endgroup

The preceding paragraphs describe these papers in terms of their later influence on research agendas. Our rubric classifies them more narrowly at the level of the text itself: most are \emph{descriptive} or \emph{reformist}, and none are \emph{normative}. This distinction matters. The F1 axis measures the form of a paper's stated claim, not its eventual agenda-setting effect. The comparison below therefore relies on \texttt{artifact\_type}, where the coding is stable under both interpretations.

Across these cases, the recurring pattern is that agenda-shifting papers bundled their claims with a \emph{new load-bearing artifact}---not necessarily a released model, but an operationalizing object such as a measurement protocol, benchmark proposal, toy implementation, dataset card, audit template, or falsifiable experimental program---rather than only with a critique of an existing benchmark or evaluation. This is not a statistical claim about all agenda-shifting work; critique-of-existing papers can and do shift agendas, and our sample is small and hand-selected. The claim is compositional: this mode of coupling is weakly represented in the accessible 2025 pool, where the second-most common artifact mode ($44/191 = 23\%$) is critique of an existing benchmark, released datasets are rare after re-reading ($3/191 = 1.6\%$), and the apparent \texttt{released\_model} case is a signing library rather than model weights. The absence of model releases is not the prescription; it is one symptom of a broader scarcity of submissions that make a new direction operational.

Taken together with F3, this locates the gap more precisely. The historical mode of coupling is scarce in the accessible pool, and we find no evidence that reviewer ratings track artifact coupling when it appears. Table~\ref{tab:artifact-type} shows the second point: neither reviewer rating nor acceptance varies systematically by \texttt{artifact\_type}. The \texttt{released\_dataset} row ($n{=}3$) speaks to submission composition, not reviewer preference. The gap is compositional, not evaluative.

\begin{table}[t]
\centering\small
\caption{Post-hoc: reviewer rating and acceptance by \texttt{artifact\_type} (pooled $N = 191$; ICML is normalized ratings on 1-10). Neither mean rating nor acceptance rate varies systematically across artifact types.}
\label{tab:artifact-type}
\begin{tabular}{lrrrr}
\toprule
\texttt{artifact\_type} & $n$ & Mean rating & SD & Accept rate \\
\midrule
existing\_benchmark\_critique & 44 & 6.02 & 1.54 & 65.9\% \\
measurement\_study            & 88 & 6.40 & 1.06 & 56.8\% \\
proposed\_experiment          & 32 & 5.91 & 1.55 & 62.5\% \\
none                          & 24 & 6.51 & 1.24 & 50.0\% \\
released\_dataset             &  3 & 6.56 & 0.51 & 66.7\% \\
\bottomrule
\end{tabular}
\end{table}

\section{Mechanisms: Why the 2025 Reviewed Pool Is Dominated by Reformist Critique}
\label{sec:mechanism}

We propose five mechanisms that help explain the dominance of reformist critique in the publicly accessible 2025 reviewed pool. The first four are local incentives on authors, reviewers, and career structure. The fifth is the aggregate signal those incentives can produce.

\paragraph{M1. Reformist claims are locally falsifiable; normative claims are globally hard to adjudicate.} A reformist claim of the form ``benchmark $X$ overfits to artifact $Y$'' can be checked: run the experiment. A normative claim of the form ``the field should prioritize $Z$'' requires the reviewer to make a judgment about the long-run trajectory of research, which is intrinsically subjective and therefore harder to defend in a written review. Reviewers under workload pressure will systematically prefer the former, creating an acceptance gradient that authors anticipate.

\paragraph{M2. The field's evidentiary culture rewards empirical rigor by default, and reformist claims admit easier empirical support.} ML has a strong culture of demanding evidence: a self-image as an experimental discipline visible in the field's heavy weighting of benchmark numbers, ablations, and error bars in evaluative judgments. A reformist critique of an existing practice can produce crisp measurements; a normative proposal for a new direction often cannot (the direction does not yet exist to be measured). The rubric we hypothesized reviewers would use, artifact coupling, turned out not to be what they check (F3), but the underlying cultural pressure is the same: papers with more immediate, grounded empirical handles get written because they are easier to publish.

\paragraph{M3. Career incentives favor reformist claims.} A reformist paper generates a discrete, citable contribution that slots neatly into the author's publication record. A genuinely normative paper (one proposing a direction not yet pursued) risks being seen as speculation by hiring committees, grant reviewers, and tenure committees. Junior researchers face asymmetric risk in the normative direction. The current track design does little to offset this asymmetry.

\paragraph{M4. The track's review criteria under-specify what good agenda-setting looks like.} The current CFP asks reviewers to judge ``novelty, rigor, and significance'' without operationalizing what these mean for a normative or predictive claim. Rigor is legible for a reformist argument (does the critique actually hold?) but opaque for a normative one (is the proposed direction important?). In the absence of explicit criteria, reviewers default to evaluating rigor in the dimension they can see; authors learn what that is.

\paragraph{M5. Local incentives aggregate into a field-level signal about what counts as position-paper work.} M1--M4 describe pressures on individual papers at individual decision points; their consequence is aggregate. When a high-status venue's visible reviewed pool is dominated by one contribution form, junior researchers learn what a ``Position Paper'' looks like by example. The evidence here is compositional, not longitudinal: a reviewed pool with few papers centered on new operationalizing artifacts sends a different signal from a charter that names ``advancing a new perspective.''

None of these mechanisms is individually novel; their relevance is their joint fit with the observed pattern: rigor in the submissions (F2) that does not predict reviewer ratings (F3), and a difference in artifact kind from the historical reference class, with at most a trend-level ratings differential favoring ambition (F4).

\section{Implications: Four Interventions for 2026 and 2027}
\label{sec:implications}

If the audit's pattern is approximately correct, the Position Paper Track's criteria need not be broken for its visible submission mix to be narrow. The substantive target of our interventions is \emph{artifact mode} (operationalization-of-new versus critique-of-existing), the axis along which we identify a contrast between the accessible reviewed pool and the historical reference class. Interventions 1 and 3 use claim type as an administrative proxy (it is easier for an author to self-report and for a reviewer to apply than artifact mode); Intervention 4 names artifact mode directly. We order them from lightest to heaviest.

\subsection{Intervention 1: Revise the submission prompt to ask claim type explicitly}

The 2026 CFP already asks authors for a written rationale explaining why the submission is suited to the Position Paper Track. We suggest adding a single required field: \emph{``What is the primary claim type of this paper? (a) normative: propose a new direction; (b) predictive: claim about what will matter; (c) reformist: propose a change to current practice; (d) descriptive: characterize the state of the field.''} This field would be visible to reviewers but not used as an acceptance filter. Its purpose is signaling: it forces authors to articulate the ambition of their contribution, and it gives reviewers a category against which to evaluate fit.

The intervention is cheap, reversible, and leaves a dataset for the 2027 chairs: if self-reported claim types match the LLM-audited distribution, the pattern is confirmed from authors' own accounts; if not, the chairs learn something about the gap between paper positioning and paper execution.

\subsection{Intervention 2: Add an explicit falsifiability sentence to the track template}

Our audit found that only 0.5\% of submissions contain an explicit falsifiability sentence, though the classifier judged 100\% of papers to contain an inferable one. We suggest requiring a one-sentence ``Falsifiability statement'' at the end of the introduction, of the form: \emph{``The central claim of this paper would be undermined by the following observation or experiment: \_\_\_.''} Missing or vacuous completions would be grounds for revision, not desk rejection.

The effect is to convert a norm currently held implicitly into one enforced explicitly: near-zero cost to serious authors, substantial cost to submissions whose claims turn out to be unfalsifiable under forcing.

\subsection{Intervention 3: Create distinct review criteria for normative/predictive vs. reformist papers}

To be clear about what this intervention is not: we are not proposing that the track discourage reformist submissions. Reformist papers have a real role, and the community loses something if it cannot publish rigorous critiques of existing practice. What we want to prevent is the track's \emph{implicit} sorting becoming its \emph{de facto} definition: position papers becoming synonymous with reformist critiques, while normative and predictive contributions move to blog posts, keynotes, and panel discussions.

The root cause of this compositional imbalance, on our reading, is that ``rigor'' means different things for different claim types. For a reformist paper, rigor means demonstrating that the critiqued practice actually fails; for a normative paper, it means demonstrating that the proposed direction is neither obviously wrong nor already being pursued. The current CFP bundles these, favoring the legible form. We suggest splitting review criteria along the claim-type axis:

\begin{itemize}[leftmargin=1.4em, topsep=1pt, itemsep=0pt]
\item For reformist papers: does the critique hold? Is the proposed alternative actionable? Does the paper quantify the stakes of the current practice's failure?
\item For normative/predictive papers: is the proposed direction sufficiently specified to be pursued? Is it not already a mainstream research agenda? Does the paper articulate what evidence, if it arrived, would confirm or undermine the position?
\end{itemize}

Reviewers should be asked which criteria they applied, and the chairs should track the distribution.

\subsection{Intervention 4: Target the submission mix}

The bluntest intervention: the chairs could explicitly solicit direction-setting submissions in the CFP language, using two framings that the historical reference class suggests are complementary. The first is claim-type-level: papers making normative or predictive claims about the field. The second is artifact-mode-level: papers that bundle a position with a \emph{new} operationalizing artifact rather than only a critique of an existing one, as a direct response to the reference-class pattern. A CFP that names the imbalance on both axes (``we welcome reformist critiques but are particularly interested in papers that propose new directions, make falsifiable predictions, or bundle their position with a new operationalizing artifact'') would cost nothing and might shift the mix over one or two cycles.

This intervention asks the track to state a substantive editorial preference, not merely an evaluative criterion. The visible reviewed pool already sends such a signal by default; the honest move is to make the choice deliberate rather than emergent.

\section{Limitations}
\label{sec:limitations}

We take our own empirical limitations seriously because our paper is \emph{about} the evidentiary standards of a research track. Six limitations matter, in descending order of importance.

\paragraph{L1. LLM classifier noise.} The artifact-coupling score showed only 0.50 cross-model agreement between Gemini and Claude. This is the field central to F3. F3 is robust to this: the pooled $\rho = +0.042$ is close enough to zero that classifier noise could not move it to a meaningful correlation; the null holds per-venue ($\rho = -0.028$ NeurIPS, $\rho = +0.067$ ICML); and Gemini's 3-run self-consistency on this field is 0.85, so the disagreement is between models, not within. Claim-type classification had 0.85 cross-model agreement: acceptable but not razor-sharp, and likely an upper bound because the Claude re-audit saw Gemini's output. The blinded human re-audit in \S\ref{sec:audit} (0.81 raw agreement) supplies a cleaner protocol; Gemini's 3-run self-consistency on claim\_type (0.90) is the tightest anchor-free reliability floor. The qualitative F1 pattern (reformist majority, normative minority) survives re-attribution of the 15\% disagreement rate unless that disagreement is both systematic and directional.

\paragraph{L2. F4 is post-hoc and does not survive scale-correct pooling.} As reported in F4 (\S\ref{sec:audit}), the pooled normative-vs-reformist differential is null under ICML$\times 2$ normalization; the raw-pooled $p = 0.034$ effect we identify there is a scale-mixing aggregation artifact. What remains is a small within-venue trend, consistent in direction but modest in magnitude. A reviewer who treats F4 as unsupported has a legitimate basis for doing so; the paper's argument rests on F1, F3, F5, and the historical reference class, none of which depend on F4.

\paragraph{L3. Venue selection bias.} We audited NeurIPS 2025 and ICML 2025, the two largest ML-specific position-paper venues. The reformist-dominance pattern may still be venue-specific; these data cannot rule that out.

\paragraph{L4. Reviewer score as outcome measure.} We use the average reviewer rating as the dependent variable in F3 and F4. Acceptance is a richer signal but is complicated across venues by the sample-completeness issue (withdrawn rejections are not public on OpenReview, so within-pool accept/reject proportions do not reflect venue acceptance rates) and by the scale difference addressed in \S\ref{sec:audit}. Using within-corpus acceptance as a robustness check, F3 stays null ($p = 0.72$) and F4's within-venue direction persists (63.6\% accept rate for normative vs.\ 59.4\% reformist). Conclusions qualitatively survive either choice.

\paragraph{L5. Cross-sectional evidence, not longitudinal.} Our audit covers two venues (NeurIPS 2025, ICML 2025) at one point in time. We do not claim to observe longitudinal change. The primary descriptive claim is that the publicly accessible 2025 reviewed pool is dominated by reformist critique, when read against the track's stated charter and the historical reference class in \S\ref{sec:history}. The substantive prescription is unchanged: if the charter names direction-setting as the track's purpose, the 74.9\%/0\% composition is a gap worth closing, whether or not it is temporally novel.

\paragraph{L6. Claim type is a weak proxy for direction-setting.} The axis F1 measures is not the axis our argument ultimately cares about: the substantive target is artifact mode (operationalization-of-new vs.\ critique-of-existing), not claim type. We use claim type in Interventions 1 and 3 as an administrative proxy because it is easier to self-report and apply than artifact mode; Intervention 4 names artifact mode directly. If a future audit finds the proxy does not track artifact mode, the correct response is to replace the proxy, not abandon the rebalancing goal.

\section{Alternative Views}
\label{sec:alternative}

We engage three substantive objections to our prescription. Methodological concerns about the audit belong in \S\ref{sec:limitations}; here the question is whether the track should \emph{not} rebalance.

\paragraph{Alternative 1: Reformism is a prerequisite, not a competitor, to direction-setting.} A proponent of this view accepts our F1 but reads it oppositely: you cannot set a new direction without first clearing the ground. Critique of existing practice is the precondition for agenda-setting work, not its alternative. Our own historical reference class confirms this, not contradicts it: before the Transformer could displace recurrence, the field had to see recurrence's failure modes; Stochastic Parrots' warnings landed because the LLM paradigm had already been critiqued piecemeal. A venue that hosts rigorous critique is the venue where the next direction-setting paper will find its target. On this reading, the 74.9\% reformist plurality is not a problem but a maturation: the track is healthy because it is identifying where new directions are needed, and the direction-setting papers will follow naturally.

We accept this reading in part. Reformism and direction-setting are not enemies; the historical examples often sit atop earlier critiques. Our disagreement is narrower: the view assumes direction-setting will follow automatically, whereas our evidence suggests it may require explicit solicitation. The 74.9\% reformist plurality, combined with the scarcity of new operationalizing artifacts, suggests the prerequisite is present in abundance while the downstream mode remains undersolicited.

\paragraph{Alternative 2: The reformist/direction-setting distinction is artificial; papers do both at once.} On this reading, a well-constructed position paper typically critiques an existing practice and proposes a new direction in the same move. ``Benchmark $X$ is saturated; we should adopt $Y$'' is simultaneously reformist (in its critique of $X$) and normative (in its proposal of $Y$); forcing it into a single category inflates the apparent reformist plurality and fabricates a gap that does not exist.

We grant the surface observation: many papers in the pool bundle critique with proposal, and our primary-orientation coding compresses that. But our central contrast is artifact type, not claim type. A paper can be reformist and normative in its prose, yet still reveal whether its position is operationalized through something new or through a critique of an existing object. Of 191 submissions, the pool rarely centers a new operationalizing artifact of the kinds we identify. The claim-type ambiguity this view identifies is real; the artifact-mode gap survives it.

\paragraph{Alternative 3: ML is an engineering discipline and reformist dominance is a feature, not a bug.} The stronger version of this view rejects our framing entirely: machine learning's progress comes through iterative refinement of artifacts and evaluation, the community is correctly self-selecting toward contributions that can be built on, and the historical reference class we cite is atypical: the exceptions rather than the rule for what has moved ML. On this view, a venue optimized for direction-setting would not produce more Concrete Problems papers; it would produce more speculative essays the community cannot act on.

Much of this we take as given. ML's engineering culture is real; F2 shows the community has absorbed the engineering standard; we would not defend a track that accepted primarily grand proclamations with no empirical handles. What we reject is the claim that this engineering view rules out direction-setting as a legitimate mode. The historical reference class shows that ML's engineering discipline has historically proceeded through \emph{both} iterative refinement and occasional direction-setting interventions: the interventions being not speculative essays but papers that bundled a claim with a new operationalizing artifact, which can include a measurement protocol, benchmark proposal, toy implementation, dataset card, audit template, or falsifiable experimental program. A track that hosts both is consistent with the engineering view as long as direction-setting contributions meet the same evidentiary bar. Our interventions are designed to enforce that bar (Intervention 3) while making room for the mode (Intervention 4).

\vspace{-0.2cm}
\section{Conclusion}
\label{sec:conclusion}

The Position Paper Track was created to host papers that advance new perspectives and propose new research directions. Its publicly accessible 2025 reviewed pool shows a narrower visible equilibrium. Reformist critiques dominate the pool, outnumbering all other claim types combined by nearly 3:1, while papers that make new directions operational through measurement protocols, benchmark proposals, toy implementations, dataset cards, audit templates, or falsifiable experimental programs are comparatively rare.
This pattern should not be read as a failure of the papers the track received. Many reformist critiques are rigorous, useful, and well-suited to a position venue. The concern is instead one of venue formation. In a new track, the reviewed papers that become publicly visible help teach future authors what kinds of positions are legible, what forms of evidence are reviewable, and which kinds of ambition can survive peer review. If the visible pool is dominated by locally falsifiable critiques of existing practice, then the track may gradually narrow the very category it was created to broaden.

Our evidence does not show that reviewers reject ambitious proposals outright, nor that reformist critique is overvalued in any simple sense. Review scores show no reliable sensitivity to the evidentiary depth captured by our rubric, and the weak ratings advantage for ambition is too fragile to carry the argument. The stronger evidence is compositional: what becomes visible in the reviewed pool, and what kinds of artifacts those papers ask the community to build on, test against, or contest.

Our proposed interventions are accordingly modest. They are not remedies for the deeper structural forces that make some kinds of agenda-setting work easier to produce, resource, and review than others. They are CFP-level nudges: clarify that operationalizing artifacts can be lightweight; ask authors to state what subsequent work their position enables; make alternative views a site of genuine contestation; and signal that credible position papers may propose research directions as well as critique existing ones. Such changes are small, but timing matters. While the track is still forming, small signals can shape the equilibrium it settles into.

\begin{tcolorbox}[
  colback=red!3,
  colframe=red!70!black,
  boxrule=0.6pt,
  arc=2pt,
  left=8pt, right=8pt, top=8pt, bottom=0pt,
  title=\textbf{Takeaway},
  coltitle=white,
  colbacktitle=red!70!black,
  attach boxed title to top left={xshift=10pt, yshift=-8pt},
  boxed title style={arc=2pt, boxrule=0pt},
  enhanced
]
\textbf{Make Positions Operational:}\ \emph{A position paper shifts a research agenda when it gives others something concrete to build on, test against, or contest.}
\end{tcolorbox}

\bibliographystyle{unsrtnat}
\bibliography{references}

\clearpage
\appendix

\section{Classification Rubric, Pipeline, and Validation Protocol}
\label{app:rubric}

This appendix documents the classification instrument used for the audit in \S\ref{sec:audit}. A reviewer or future replicator should be able to reconstruct the classification task from the text below, independent of our code release.

\subsection{Rubric prompt (verbatim)}
\label{app:rubric-prompt}
 
The following text was passed verbatim to Gemini 3 Pro for every paper in the corpus. A single placeholder \texttt{\{paper\_text\}} was replaced with the extracted paper body (truncated to a maximum character length set in our pipeline configuration). The prompt was unchanged across all 191 papers and across all three independent runs per paper.
 
\begin{quote}\small
\emph{You are auditing a Position Paper from an ML conference. Read the paper text and return only valid JSON matching the schema exactly --- no prose, no code fences, no extra keys. Every field must be filled with a non-null value chosen from the allowed values listed below.}
 
\medskip
\emph{Schema (fill in every field; do not output the placeholders):}
 
{\footnotesize
\begin{verbatim}
{
  "thesis_one_sentence": "<one-sentence summary of the paper's thesis>",
  "claim_type": "<one of: normative | predictive | descriptive | reformist>",
  "artifact_coupling_score": "<integer 0-5>",
  "artifact_type": "<one of: none | existing_benchmark_critique
                    | proposed_experiment | measurement_study
                    | released_dataset | released_model>",
  "falsifiability_statement_present": "<true | false>",
  "falsifiability_inferred": "<true | false>",
  "target_audience": "<one of: ml_researchers | policy | public | mixed>",
  "cites_any_benchmark_data": "<true | false>"
}
\end{verbatim}
}
 
\emph{\textbf{claim\_type:}}
\begin{itemize}
\item \emph{normative --- the paper argues the community ought/should pursue a new direction}
\item \emph{predictive --- the paper makes a falsifiable claim about future empirical developments}
\item \emph{descriptive --- the paper characterizes the current state of the field without an ought-claim}
\item \emph{reformist --- the paper proposes a specific method, practice, or institutional change}
\end{itemize}
\emph{Choose the paper's PRIMARY orientation. If genuinely balanced between two, pick the one that the thesis sentence leans toward.}
 
\emph{\textbf{artifact\_coupling\_score (0--5):}}
\begin{itemize}
\item \emph{0 --- purely argumentative; no empirical material of any kind}
\item \emph{1 --- references existing results in passing to motivate a claim}
\item \emph{2 --- re-analyzes or non-trivially critiques a specific existing benchmark / dataset}
\item \emph{3 --- proposes a concrete, runnable experiment or measurement that would falsify or support the thesis, but does not execute it}
\item \emph{4 --- executes a measurement study, small experiment, or quantitative audit whose outcome is load-bearing for the thesis}
\item \emph{5 --- releases a dataset, benchmark, model, or tool that operationalizes the position so others can build on or contest it}
\end{itemize}
 
\emph{\textbf{artifact\_type:}}
\begin{itemize}
\item \emph{none --- no identifiable artifact}
\item \emph{existing\_benchmark\_critique --- primary contribution is critique / re-analysis of an existing benchmark or dataset}
\item \emph{proposed\_experiment --- proposes a concrete experiment / measurement but does not execute it}
\item \emph{measurement\_study --- executes a measurement study, audit, or empirical survey}
\item \emph{released\_dataset --- releases a new dataset as the primary artifact}
\item \emph{released\_model --- releases new model weights as the primary artifact}
\end{itemize}
\emph{If the paper has multiple artifacts, pick the one most load-bearing for the thesis.}
 
\emph{\textbf{falsifiability\_statement\_present:} true ONLY if the paper contains an explicit passage stating what observation or experiment would refute its thesis. A vague gesture toward ``future work'' does not count.}
 
\emph{\textbf{falsifiability\_inferred:} true if you (the classifier) can articulate at least one plausible refuting observation from the paper's claims, whether or not the paper itself states one.}
 
\emph{\textbf{target\_audience:}}
\begin{itemize}
\item \emph{ml\_researchers --- primarily persuading the ML research community}
\item \emph{policy --- primarily persuading policymakers or regulators}
\item \emph{public --- primarily persuading a general audience}
\item \emph{mixed --- no single audience dominates}
\end{itemize}
 
\emph{\textbf{cites\_any\_benchmark\_data:} true if at least one numerical result from a benchmark or dataset is cited anywhere in the main body; false otherwise.}
 
\medskip
\emph{Paper text:}
 
\emph{\texttt{\{paper\_text\}}}
\end{quote}

\subsection{Reading the rubric as deployed}
\label{app:rubric-interp}

A few operational points clarify how the rubric behaves on real papers, and matter for interpreting the audit's findings.

\paragraph{Claim-type boundary.} The rubric distinguishes \emph{normative} (ought/should claims about the future direction of the field) from \emph{reformist} (proposes a specific method or institutional change). In practice, the distinction is: normative claims are about \emph{where the field should go}; reformist claims are about \emph{how an existing practice should be changed}. A paper arguing ``the community should invest in scaling laws for X'' is normative; a paper arguing ``benchmark Y is saturated and we should replace it with Z'' is reformist. The 0.85 cross-model agreement reflects the fact that most papers have a clearly primary orientation, with ambiguous cases concentrated near the boundary between broad reformism and narrow normativity.

\paragraph{Artifact-coupling score.} The 0--5 scale rewards \emph{how tightly the paper's thesis is bound to new empirical material}. A paper at level 4 (the modal score) executes a measurement study whose results are load-bearing; at level 2, it re-analyzes or critiques an existing benchmark without new data. The 0.50 cross-model agreement on this field---our lowest per-field agreement---reflects the subjectivity of deciding whether a study is ``load-bearing'' or merely ``illustrative'' and whether a critique is ``non-trivial'' or ``passing.'' We treat this as a warning against over-interpreting fine score differences and rely primarily on the coarser observation that 80\%+ of papers score $\geq 2$.

\paragraph{Artifact-type categories.} The six categories are intended to be mutually exclusive. The classifier was instructed to pick the \emph{primary} artifact orientation; papers that release both a dataset and run a measurement study (the empirically common case) are coded to whichever is more load-bearing for the thesis. This coding choice partly explains the 0.60 cross-model agreement on this field.

\paragraph{Falsifiability indicators.} The two indicators intentionally measure different things. The first (``statement present'') is a strict textual test: does the paper explicitly say what would refute it? The second (``inferred'') is weaker: could the classifier reasonably articulate a refutation from the paper's claims? The 0.5\% vs.\ 100\% gap we report in F5 reflects this distinction and motivates Intervention 2.

\subsection{Pipeline}
\label{app:pipeline}

The end-to-end audit pipeline consists of the following steps (implemented as separate Python scripts; code will be released with the final version of this paper):

\begin{enumerate}
\item \textbf{Corpus collection.} OpenReview API pull of all Position Paper Track submissions from NeurIPS 2025 and ICML 2025. Filtered to submissions with a retrievable PDF and at least one completed review. Final $N = 191$.
\item \textbf{Text extraction.} PDF text extraction via standard Python tooling (\texttt{pdfplumber}). Extracted text stored per paper as plain text.
\item \textbf{Classification.} Each paper classified 3 times independently by Gemini 3 Pro using the verbatim rubric in \S\ref{app:rubric-prompt}. Papers with fewer than 2 successful runs (parsing failures, API errors) were dropped; all 191 final papers had at least 2 successful runs, and 99.5\% had all 3.
\item \textbf{Consolidation.} For each field, the consolidated label is the majority vote across the 3 runs. Ties broken toward the lexicographically smaller value (for sortable types) or arbitrarily (for non-sortable). We also record a per-field, per-paper self-agreement score (fraction of 3 runs matching the majority).
\item \textbf{Cross-model validation.} All 191 papers were re-classified by an independent LLM (Claude Sonnet 4.6). Cross-model agreement is reported per field as the fraction of papers on which Claude's label exactly matched Gemini's consolidated label.
\item \textbf{Analysis.} Statistical tests (Spearman, Mann-Whitney, Kruskal-Wallis, Fisher exact, pairwise with Bonferroni) computed over the consolidated labels joined to OpenReview review scores and decisions.
\end{enumerate}

\subsection{Validation protocol and an honest caveat}
\label{app:validation}

The validation covers the full corpus ($N = 191$). Agreement was computed per field as exact-match fraction, including for the 0--5 integer coupling score.

\textbf{Anchoring caveat.} When Claude Sonnet 4.6 was asked to re-classify each paper, the Gemini consolidated output for that paper was included in the task context (as part of the per-paper record, alongside the paper text). We did not instruct Claude to blind itself to the Gemini output before forming its own classification. As a result, the 0.85 claim-type agreement and other cross-model agreement numbers we report should be treated as \emph{upper bounds} on the true cross-model concordance: a cleaner blind protocol would likely show lower agreement.

We report the numbers as observed rather than adjusting them downward. The Gemini 3-run self-consistency scores (artifact-coupling 0.85, claim-type 0.90, artifact-type 0.84, others $\geq$ 0.93) are not subject to this anchoring concern; they are tabulated in Table~\ref{tab:reliability} and provide the cleaner reliability signal for the fields in question.

\section{Additional Analyses}
\label{app:additional}

This appendix reports supporting analyses that are referenced in the main text but exceed its space budget. All tests here are post-hoc (not pre-specified in H-A1--H-A4) and are reported for transparency rather than as primary evidence.

\subsection{Pre-specified hypothesis tests in full}
\label{app:ha-table}

Table~\ref{tab:ha-full} reports the four pre-specified tests per venue and pooled. H-A1 (Spearman) and H-A2 (Mann-Whitney on artifact-coupling by accept/reject) are both null pooled (H-A1 pooled uses the ICML$\times 2$-normalized rating, per \S\ref{sec:audit}); H-A4 (sign consistency across venues) fails on both H-A1 and H-A2; H-A3 (Fisher exact on \texttt{falsifiability\_inferred} $\times$ acceptance) is degenerate because the classifier inferred falsifiability for \emph{every} paper in the corpus (100\%), leaving no variance for the test.

\begin{table}[h]
\centering\small
\caption{Pre-specified hypothesis tests. H-A1: Spearman $\rho$ between artifact-coupling score and reviewer rating (pooled row uses ICML$\times 2$-normalized rating; per-venue rows are rank-invariant under the rescaling). H-A2: Mann-Whitney U on artifact-coupling by accept/reject (scale-invariant). H-A4: sign consistency of H-A1/H-A2 across venues. H-A3 omitted from table: degenerate ($100\%$ of papers have \texttt{falsifiability\_inferred = True}).}
\label{tab:ha-full}
\begin{tabular}{lrrrrrl}
\toprule
Split & $N$ & H-A1 $\rho$ & H-A1 $p$ & H-A2 $U$ & H-A2 $p$ & H-A2 direction \\
\midrule
Pooled   & 191 & $+0.042$ & 0.57 & 4533.5 & 0.72 & accept higher \\
NeurIPS  & 95  & $-0.028$ & 0.79 & 938.5  & 0.20 & reject higher \\
ICML     & 96  & $+0.067$ & 0.51 & 1027.0 & 0.09 & accept higher \\
\midrule
\multicolumn{7}{l}{H-A4 sign-consistent across venues: \textbf{false} on both H-A1 and H-A2.} \\
\bottomrule
\end{tabular}
\end{table}

\subsection{Artifact-type composition by venue}
\label{app:artifact-type-venue}

Table~\ref{tab:artifact-type-venue} reports the \texttt{artifact\_type} distribution split by venue, with the same two classifier labels reassigned after manual re-reading (1 ICML \texttt{released\_dataset} $\rightarrow$ \texttt{measurement\_study}; 1 ICML \texttt{released\_model} $\rightarrow$ \texttt{proposed\_experiment}) as in Table~\ref{tab:artifact-type}. The artifact-kind contrast identified in \S\ref{sec:history} holds in both NeurIPS and ICML independently in the weaker sense that new primary artifacts are rare: genuine released-dataset submissions are rare and present at both venues (2/95 NeurIPS, 1/96 ICML), the apparent released-model case does not release model weights, and \texttt{existing\_benchmark\_critique} is the second-most common artifact mode at both (19/95 NeurIPS, 25/96 ICML). The compositional signal is not an artifact of venue aggregation.

\begin{table}[h]
\centering\small
\caption{\texttt{artifact\_type} composition by venue, after reassigning two of the five schema-nominal \texttt{released\_dataset} / \texttt{released\_model} papers on manual re-reading (1 ICML \texttt{released\_dataset} $\rightarrow$ \texttt{measurement\_study}; 1 ICML \texttt{released\_model} $\rightarrow$ \texttt{proposed\_experiment}); the remaining three \texttt{released\_dataset} papers (2 NeurIPS, 1 ICML) are retained. Columns sum to 95 (NeurIPS) and 96 (ICML). Both venues exhibit the same qualitative pattern: dominant \texttt{measurement\_study}, substantial \texttt{existing\_benchmark\_critique}, rare dataset releases, and no genuine model-weight releases.}
\label{tab:artifact-type-venue}
\begin{tabular}{lrrrr}
\toprule
\texttt{artifact\_type}        & NeurIPS $n$ & NeurIPS \% & ICML $n$ & ICML \% \\
\midrule
measurement\_study             & 43 & 45.3\% & 45 & 46.9\% \\
existing\_benchmark\_critique  & 19 & 20.0\% & 25 & 26.0\% \\
proposed\_experiment           & 17 & 17.9\% & 15 & 15.6\% \\
none                           & 14 & 14.7\% & 10 & 10.4\% \\
released\_dataset              &  2 &  2.1\% &  1 &  1.0\% \\
\bottomrule
\end{tabular}
\end{table}

The pattern shown in main-text Table~\ref{tab:artifact-type} is post-hoc; \texttt{artifact\_type} had the second-lowest cross-model agreement in validation (0.60), so we read the absence of systematic variation across artifact types as evidence consistent with F3 rather than a definitive claim about reviewer behavior.

\subsection{Acceptance rate by claim type}
\label{app:claimtype-accept}

Table~\ref{tab:claimtype-accept} reports within-pool acceptance rates by claim type. The normative-vs-reformist gap (63.6\% vs.\ 59.4\%) goes in the same direction as the within-venue rating-based trend reported in F4 (normative slightly above reformist at each venue), though acceptance is a noisier outcome (L4 in \S\ref{sec:limitations}) and the acceptance ordering of the two low-$n$ cells (predictive $n=4$, descriptive $n=11$) should not be read as an effect. No column of this table survives a significance test; we report it for transparency only.

\begin{table}[h]
\centering\small
\caption{Within-pool acceptance rate by claim type (pooled $N = 191$).}
\label{tab:claimtype-accept}
\begin{tabular}{lrrr}
\toprule
Claim type   & $n$ & $n_{\text{accept}}$ & Accept rate \\
\midrule
normative    & 33  & 21 & 63.6\% \\
reformist    & 143 & 85 & 59.4\% \\
predictive   & 4   & 2  & 50.0\% \\
descriptive  & 11  & 5  & 45.5\% \\
\bottomrule
\end{tabular}
\end{table}

\subsection{Exploratory ordinal Spearman on claim type}
\label{app:claimtype-ordinal}

As an additional exploratory check on F4, treating claim type as an ordinal variable in the order [\textsc{normative}, \textsc{descriptive}, \textsc{predictive}, \textsc{reformist}] and correlating against ICML$\times 2$-normalized reviewer rating gives Spearman $\rho = +0.010$, $p = 0.89$, $N = 191$: essentially zero. A raw-pooled version of the same test (concatenating the 1--10 NeurIPS and 1--5 ICML scales without normalization) would show $\rho = -0.12$, $p = 0.10$, but that is a scale-mixing artifact of the same kind discussed in F4. No claim in the paper rests on this test.

\subsection{Rubric classification of the historical reference class}
\label{app:historical-rubric}

For completeness we ran the same Gemini 3 Pro rubric (3 runs per paper, majority vote, identical pipeline to the 191-paper audit) on eight canonical reference-class papers named in \S\ref{sec:history}: AlexNet \citep{krizhevsky2012imagenet}, the Transformer \citep{vaswani2017attention}, the two adversarial-example papers \citep{szegedy2014intriguing,goodfellow2015explaining}, Concrete Problems \citep{amodei2016concrete}, the Bitter Lesson \citep{sutton2019bitter}, and the two RLHF papers \citep{christiano2017deep,ouyang2022training}. Stochastic Parrots \citep{bender2021stochastic} is named in \S\ref{sec:history} but was not re-classified, because it is a modern reference-class paper whose claim structure is closer to the track submissions we audit than to the engineering-paper comparand we use it against.

Table~\ref{tab:historical-rubric} reports the per-paper labels. The quantity that matters for the compositional contrast in \S\ref{sec:history} is \texttt{artifact\_type}: the reference-class papers often make their claims operational through a new model, measurement, or proposed experiment, while the 2025 Position Track audit is dominated by \texttt{measurement\_study} and \texttt{existing\_benchmark\_critique}, with only $3/191$ released datasets and no genuine model-weight releases. We do not take this as an argument that position papers should release models; rather, it motivates the weaker intervention that position papers can operationalize new directions through measurement protocols, benchmark proposals, toy implementations, dataset cards, audit templates, or falsifiable experimental programs.

The rubric classified seven of eight papers as \texttt{descriptive} or \texttt{reformist} rather than as \texttt{normative}, disagreeing with the authors' framing in \S\ref{sec:history}. This is consistent with the rubric's definition: \texttt{claim\_type} captures the text-level primary orientation of the paper as submitted, not its downstream normative effect on the field. The historical papers are engineering contributions whose normative weight was delivered through capability (AlexNet, Transformer, InstructGPT) or through measurement (the two adversarial-example papers, the Bitter Lesson); the rubric correctly reads them as engineering papers. Gemini's 3-run self-consistency across the reference class is uniformly high (mean 0.94 across all fields on all eight papers; 1.00 for both \texttt{artifact\_coupling\_score} and \texttt{artifact\_type}), so the labels are stable even where they disagree with authorial framing. The LLM-vs-author disagreement therefore does not undermine the compositional argument; it reinforces that \S\ref{sec:history}'s contrast is about artifact mode, not about claim-type language.

\begin{table}[h]
\centering\small
\caption{Rubric classification of the eight historical reference-class papers (\S\ref{sec:history}), using the same Gemini 3 Pro pipeline as the 191-paper audit. ACS = \texttt{artifact\_coupling\_score}. The \texttt{artifact\_type} column (right) supplies the compositional contrast: several reference-class papers made claims operational through new models or measurements, whereas the 2025 audit is concentrated in measurement studies and critiques of existing benchmarks.}
\label{tab:historical-rubric}
\begin{tabular}{llccl}
\toprule
Paper & Authors' framing & ACS & LLM \texttt{claim\_type} & LLM \texttt{artifact\_type} \\
\midrule
AlexNet                  & normative-by-demo     & 5 & descriptive & released\_model      \\
Transformer              & normative-by-demo     & 5 & reformist   & released\_model      \\
InstructGPT              & normative-by-demo     & 5 & reformist   & released\_model      \\
RLHF (Christiano)        & normative-by-demo     & 4 & reformist   & measurement\_study   \\
Szegedy (adversarial)    & predictive            & 4 & descriptive & measurement\_study   \\
Goodfellow (adversarial) & predictive            & 4 & descriptive & measurement\_study   \\
Concrete Problems        & predictive/normative  & 3 & reformist   & proposed\_experiment \\
Bitter Lesson            & descriptive/normative & 1 & descriptive & none                 \\
\bottomrule
\end{tabular}
\end{table}

\subsection{Full claim-type $\times$ rating pairwise comparisons}
\label{app:claimtype-full}

F4 in the main text reports the normative-vs-reformist pairwise test under ICML$\times 2$ normalization. Table~\ref{tab:claimtype-full} reports the full six-pair matrix on the normalized scale. None is significant even uncorrected. We include the full set so that readers can see the F4 pairwise was not selected from an undisclosed larger comparison set.

\begin{table}[h]
\centering\small
\caption{Post-hoc pairwise Mann-Whitney U tests on reviewer rating by claim type (pooled $N = 191$), using ICML$\times 2$-normalized ratings (\S\ref{sec:audit}). Kruskal-Wallis omnibus $H = 0.19$, $p = 0.98$ (not significant). Six pairs; Bonferroni correction multiplies each raw $p$ by 6.}
\label{tab:claimtype-full}
\begin{tabular}{lrrrrl}
\toprule
Comparison & $U$ & $p$ & $p_{\text{Bonf}}$ & $r_{\text{rb}}$ & direction \\
\midrule
normative vs.\ reformist       & 2354.5 & 0.99  & 1.0  & $0.002$ & normative higher \\
descriptive vs.\ normative     & 162.0  & 0.60  & 1.0  & 0.11    & normative higher \\
normative vs.\ predictive      &  76.0  & 0.64  & 1.0  & 0.15    & predictive higher \\
descriptive vs.\ predictive    &  20.0  & 0.84  & 1.0  & 0.09    & predictive higher \\
descriptive vs.\ reformist     & 743.0  & 0.76  & 1.0  & 0.06    & reformist higher \\
predictive vs.\ reformist      & 270.0  & 0.85  & 1.0  & 0.06    & predictive higher \\
\bottomrule
\end{tabular}
\end{table}

\subsection{Per-field classifier reliability}
\label{app:reliability}

Table~\ref{tab:reliability} reports per-field reliability from two sources: Gemini's 3-run self-consistency across all 191 papers, and cross-model agreement against Claude Sonnet 4.6 across all 191 papers (see \S\ref{app:pipeline} and \S\ref{app:validation}). The cross-model column should be read as an upper bound under the anchoring caveat in Appendix~\ref{app:validation}.

\begin{table}[h]
\centering\small
\caption{Per-field classifier reliability. Gemini self-consistency is the mean across-run agreement over 191 papers (3 runs each). Cross-model (Claude Sonnet 4.6) agreement is the exact-match fraction over all 191 papers. The cross-model figures are upper bounds because Claude saw Gemini's labels when re-classifying (Appendix~\ref{app:validation}).}
\label{tab:reliability}
\begin{tabular}{lrr}
\toprule
Field & Gemini self-consistency & Cross-model (upper bound) \\
\midrule
claim\_type                       & 0.90 & 0.85 \\
artifact\_coupling\_score         & 0.85 & 0.50 \\
artifact\_type                    & 0.84 & 0.60 \\
falsifiability\_inferred          & 1.00 & 0.95 \\
falsifiability\_statement\_present& 1.00 & 1.00 \\
target\_audience                  & 0.95 & 0.95 \\
cites\_any\_benchmark\_data       & 0.94 & 0.75 \\
\bottomrule
\end{tabular}
\end{table}

The two fields our primary findings depend on most directly---\texttt{claim\_type} (F1, F4) and \texttt{artifact\_coupling\_score} (F3)---differ in their reliability profile. Claim type has high self-consistency (0.90) and acceptable cross-model agreement (0.85 upper bound). Artifact-coupling has high self-consistency (0.85) but low cross-model agreement (0.50), which is why we rely on it only for the near-zero correlation in F3 (whose near-zero magnitude is robust to noise) and not for fine comparisons across score levels.

\paragraph{Blinded expert re-audit for \texttt{claim\_type}.} A third reliability signal for \texttt{claim\_type} comes from the blinded expert re-audit described in \S\ref{sec:audit}: a single human annotator, shown the full text of each paper, independently assigned primary claim-type labels while blinded to all model outputs, yielding raw agreement of 0.81 against the Gemini majority-vote label. Two caveats govern how much this figure can carry. First, a single annotator cannot distinguish rubric ambiguity from individual annotator idiosyncrasy; only a second independent annotator would allow inter-annotator reliability to be computed, and we did not conduct one. Second, 0.81 is raw agreement rather than a chance-corrected kappa, so the class-imbalanced distribution (74.9\% reformist) inflates the figure relative to a Cohen's-kappa-style measure. We therefore use this number only as triangulation: the blinded-human 0.81 sits slightly below the anchored-cross-model 0.85 in the direction the anchoring caveat predicts, and both sit below the anchor-free Gemini self-consistency 0.90, which remains the tightest reliability floor for this field.

\subsection{Full list of audited papers, by venue and decision}
\label{app:audited-papers}

Table~\ref{tab:audited-papers} cites all 191 audited submissions grouped by venue (rows) and within-pool decision (columns). Citation numbers are compressed using natbib's \texttt{sort\&compress} option; the underlying BibTeX entries are in the auto-generated block of \texttt{references.bib}, keyed as \texttt{or\_<paper\_id>}. The within-pool ``Rejected'' column, as discussed in \S\ref{sec:audit} and \S\ref{sec:limitations}, is not a venue-level acceptance-rate statistic because withdrawn rejections are not public on OpenReview.

\begin{table}[h]
\centering\small
\caption{All 191 audited Position Paper Track submissions grouped by venue (rows) and within-pool decision (columns). Each cell cites every paper in that group; natbib renders these as compressed numeric ranges. BibTeX entries are auto-generated in \texttt{references.bib} with keys \texttt{or\_<paper\_id>}. See \S\ref{sec:audit} for why the ``Rejected'' column is not a venue acceptance-rate statistic.}
\label{tab:audited-papers}
\begin{tabular}{lp{2.3in}p{2.0in}}
\toprule
Venue & Accepted ($n=113$) & Rejected ($n=78$) \\
\midrule
ICML 2025 ($n=96$) & \citep{or_1RlrtH6ydW, or_1ZC4RNjqzU, or_1rh8iTehBc, or_3wEY0NB2pG, or_42Au7FoD8F, or_4KhDd0Ozqe, or_4UhTWPwVke, or_5Hpm74b1Ga, or_8samCaCwKu, or_9skHxuHyM4, or_A0HtZM0MpZ, or_ACzL62Jp4E, or_Al5mEX6eHF, or_BCP8UU2BcU, or_CA9NxmmUG5, or_CYJlJgEzZs, or_DMRrbb36r5, or_ET6qJpllEi, or_ErKu9lP91g, or_GYZLed4d3M, or_GrBXso0e17, or_H72JEXAPwo, or_HuvAM5x2xG, or_IxCvgUme5S, or_J5MmGPWKfb, or_JMoWFkwnvv, or_JkcSsFWdGP, or_L6RpQ1h4Nx, or_LEYmr1TsBW, or_LL39y0Tfxb, or_LkdH35003E, or_Lrv20S5RZV, or_MxCJbuJhWG, or_Nr2ulBN50q, or_QLKBm1PaCU, or_QMgCDPWL9Y, or_RrvhbxO2hd, or_RuLsq4LSZK, or_Rxd2TpV6Eg, or_TEkyydR6il, or_UTxi86wmas, or_UXZJ3aL8vE, or_V1FP9WDKa7, or_WpePuya3Ki, or_YhZ2PY2nZa, or_YmTxiR1HUX, or_YuMEUNNpeb, or_asQJx56NqB, or_b2gM1HyAgE, or_bXfF6Dqe9s, or_cRBg1dtj7o, or_cw7MYyDL33, or_eI8KegpPyX, or_eax2ixyeQL, or_fRk0nKLKrJ, or_gCPJFcHskT, or_gnyqRarPzW, or_gwhPvu97Gm, or_hxwIndG0Z8, or_j3totqf8xW, or_j5eEF77JQz, or_jd1N60VNFE, or_kJzB6lQmcb, or_l8QemUZaIA, or_mWlnrtOTtm, or_mzc1KPkIMJ, or_nDFpl2lhoH, or_nrlGUdlo16, or_q7QJyxgAq4, or_tctWi7I5wd, or_wkisIZbntD, or_yYJo8czj4f, or_zfohnbkMu0} & \citep{or_1AS7IT3X2A, or_2qErp0n78A, or_3tdITbjv8t, or_75WZP8whT8, or_ASOb5Cw8Rv, or_DH84SmPVxR, or_ELErARGR5U, or_GCqffUAyiu, or_Gl3V36BkCL, or_I7nESnBvib, or_KL8bi77TFF, or_Njaljr4V8L, or_OT3WJRwCIf, or_QlEfwyq9rx, or_SlgXCLZFj3, or_Sz90WdONPz, or_T21ixFVruG, or_Uh13utzs2m, or_d2k8uDYlkh, or_jK4dbpEEMo, or_n1rqG1LnRF, or_vMTijVnXQ8, or_wjQqi7DJM2} \\[0.5em]
NeurIPS 2025 ($n=95$) & \citep{or_0ngi2StMwC, or_1IpHkK5Q8F, or_6plSmhBI33, or_BzFMBNqg7R, or_DS1XSAPvKs, or_Ev5xwr3vWh, or_EvXWexakZX, or_FfsxgSZW0c, or_FjxyAotxtT, or_HzGZVYi8fK, or_OMc0BYxND4, or_PFRandBfSz, or_PegEYWWXvx, or_PgA9rZoMY8, or_R5uuqCAPf8, or_RT3Jby7v21, or_SbfjBNlJE7, or_USqNoPVhxx, or_V6DcL5L6CU, or_VZnOKzQ5qW, or_a4oXTW1PW2, or_aVwhTcSMl4, or_d7hqAhLvWG, or_dl5pvd5IgW, or_fXiPp3qvrW, or_gXTFLbGUQp, or_mXBFoHDuil, or_mdKzkjY1dM, or_mfd6GRW4Az, or_qh9eGtMG4H, or_rdeCalg68L, or_tp94g4Vmad, or_uEY7kQsiZz, or_upugtLPOxC, or_vFae5rRman, or_vlM9rLv5xB, or_xcdlSMYXxD, or_xnNHXepQ9h, or_yZU8kdwafM, or_yqKfMr0yvY} & \citep{or_0TRVB5ghCR, or_4lyIMYzILk, or_5X4GDSUumr, or_816gaVGHgP, or_8Ow7kh78fk, or_8ZH52QHIZV, or_Aa50oIovvD, or_AsC0NOkJ2m, or_BXLRMWLDQw, or_EIEVNiraPS, or_FAD6MEhQQR, or_FJF1sa6elQ, or_JCqGIUAsbs, or_LAXgS0xzPf, or_NHDOjeVMb5, or_NILrMDAqEt, or_NtJfzzleG8, or_Omq9tUouSS, or_Pcys8py9RL, or_R6TXwNF1SB, or_RV12OsgCO0, or_RZRRb11jXp, or_RkX3UyGunC, or_RyBZXCVr1k, or_TDZjksboWO, or_U46jD48SJi, or_V5PNJ5HnpA, or_Vib3KtwoWs, or_XR9UpqWhmT, or_YQplP7XrAo, or_Z6UueXyYwB, or_ZOUHFrCmwu, or_a9eBWrd5Jg, or_aXMPvmBAm5, or_bEhRgt7mwt, or_cIbQaSXqYm, or_dVKcLgcCLZ, or_g8Fo6qtnMR, or_gIIqPel6w5, or_gY0BOsPO0k, or_iBkQYeEfzn, or_iOSHFKHQNP, or_j0h4glzL2F, or_j5Qmcv9jtc, or_kEdP6usKZd, or_kJfpS7lCVT, or_nKpmLCN0Q9, or_o3M9ibtZWV, or_oz2QmdrPdz, or_pRiGl7qF0v, or_tMJvb9JDsd, or_u0FB996GIH, or_uoGQOg1oxZ, or_yZhVKDW0o0, or_ygfzWIGDN8} \\
\bottomrule
\end{tabular}
\end{table}

\clearpage

\end{document}